\documentclass[a4paper,11pt]{article}
\pdfoutput=1 % if your are submitting a pdflatex (i.e. if you have
\usepackage{jinstpub} % for details on the use of the package, please
\usepackage{lineno}
\title{\boldmath The JUNO experiment and its electronics readout system}

\author[a,1]{P.-A. Petitjean,\note{Corresponding author.}}
\author[a]{B. Clerbaux,}
\author[a]{M. Collomer Molla}
\author[a]{and Y. Yang}

\affiliation[a]{Inter-University Institute For High Energies, Université libre de Bruxelles (ULB), Brussels, Belgium\\}

\emailAdd{pierre-alexandre.petitjean@ulb.be}

\abstract{The main goal of the Jiangmen Underground Neutrino Observatory (JUNO) under construction in
China is to determine the neutrino mass hierarchy and to measure oscillation parameters to the sub-percent level. The detector consists of 20 ktons of liquid
scintillators instrumented by 17612 20-inch photo-multiplier tubes, and 25600 3-inch small PMTs, with
photo-cathode coverage of 77\%. The electronics system is separated into two main parts. The front-end system, sitting under water, performs analog signal processing. The back-end electronics system,
sitting outside water, consists of the DAQ and the trigger. The design of the electronics system and the production status of experiment  will be reported in the proceedings.}


\arxivnumber{2110.12277} % only if you have one

\collaboration[c]{on behalf of JUNO collaboration}

\proceeding{
TWEPP 2021 Topical Workshop on Electronics for Particle Physics \\
  20–24 sept. 2021 \\
  Online}

\begin{document}
\maketitle
\flushbottom
\section{Introduction}
\hspace{24pt}The non zero value of the $\theta_{13}$ neutrino oscillation parameter measured by the Daya Bay \cite{Dayneut},
RENO \cite{Reno2018} and Double Chooz \cite{Chooz2016} experiments, opened the path to the determination of the Neutrino Mass Hierarchy (NMH).
The mass hierarchy measurement will give important clues for the quest of the neutrino nature (Dirac or Majorana) towards the formulation of a theory of flavour.
The NMH, Normal Hierarchy (NH) or Inverted Hierarchy (IH), has an effect on the electron anti-neutrino energy spectrum coming from the nuclear reactor. The major goal of JUNO is the determination of the NMH using electron anti-neutrinos from the nearby nuclear reactors situated at a distance of about 53 km. 
JUNO will allow for a better measurements on the other mixing parameters ($\sin^2\theta_{12}, \, \Delta m _{12}^2 ,\, \Delta m _{31}^2  \, \mathrm{and}\, \Delta m _{32}^2$ ). 
JUNO will use inverted-beta decay interactions ( $\overline{\nu_e}+p \to n+e^+$) to detect the electric anti-neutrinos coming from the fission of $^{236} \mathrm{U}$, $^{238} \mathrm{U}$, $^{239} \mathrm{Pu}$ and $^{241} \mathrm{Pu}$ inside the nuclear reactors.
The positron ($e^+$) deposits its energy and annihilates into two 511 keV $\gamma$.
The neutron, after a thermalization process, is captured by a proton around $\sim 200 \mu \mathrm{s}$ later and release a 2.2 MeV $\gamma$. 
This prompt-delayed signal spatial and temporal coincidence works as a important background suppressor. The expected daily number of reactor anti-neutrino events is around 80. In addition to the reactor anti-neutrino physics, JUNO physics program includes as well the study of solar neutrinos, supernova neutrinos, atmospheric neutrinos and Geo-neutrinos.  The requirements to perform the different physics programs are the following:
1) signal range: from 1 p.e.(photo-electron) to 100 p.e. with a linear response and
charge resolution from 0.1 p.e. to 1 p.e.; this requires a noise level below 0.1 p.e. for single p.e. detection.
2) background range: from 100 p.e. to 1000 p.e. with a resolution of 1 p.e. 3) signal rise time around 2.5 ns. This last requirement translates in a
bandwidth of about 400 MHz and therefore a sampling rate of 1 Gsample/s is appropriate\cite{Bellato2021}.

\section{Detector}
\begin{figure}
    \centering
    \includegraphics[height=6.5cm]{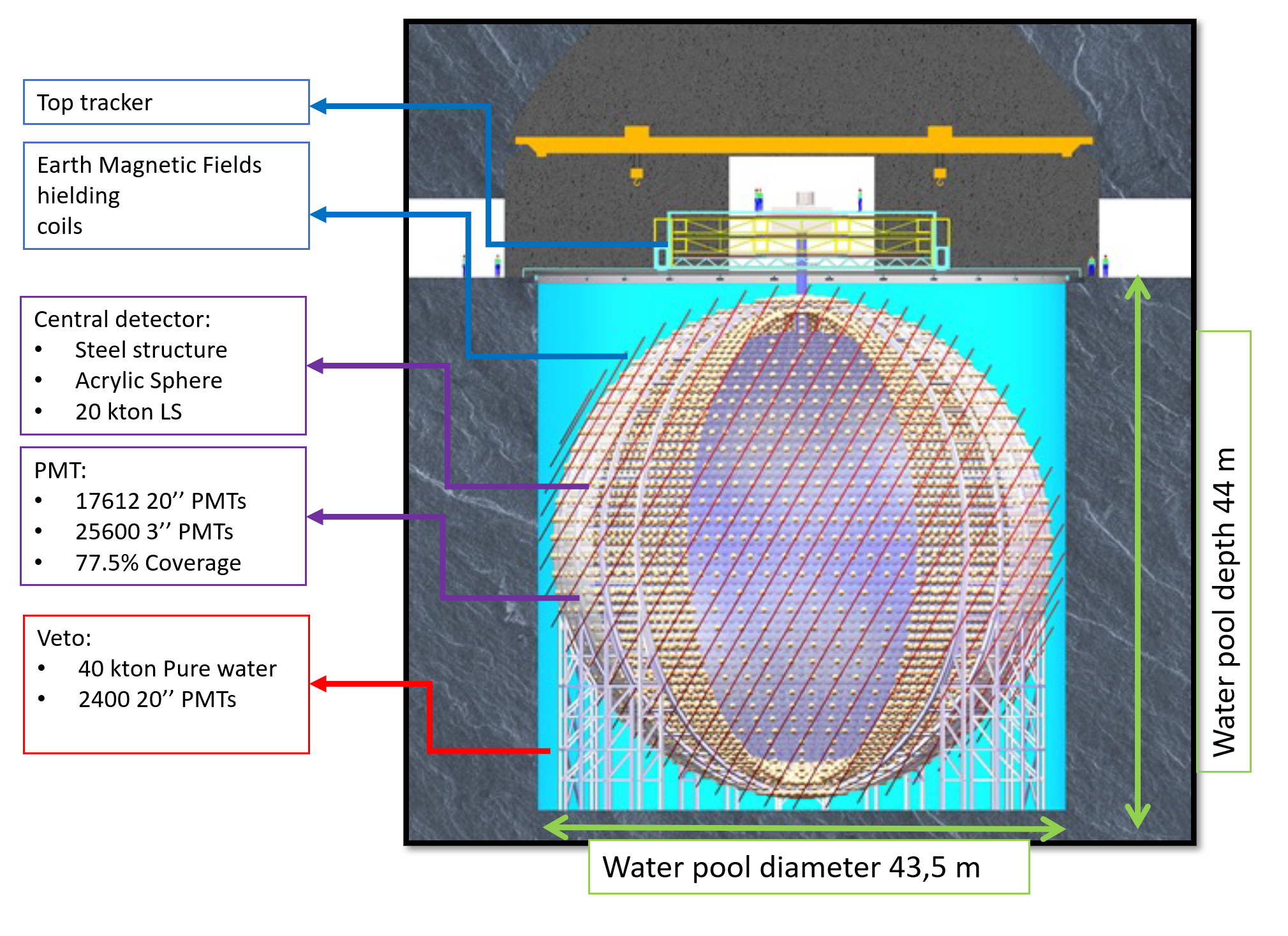}
    \includegraphics[height=6cm]{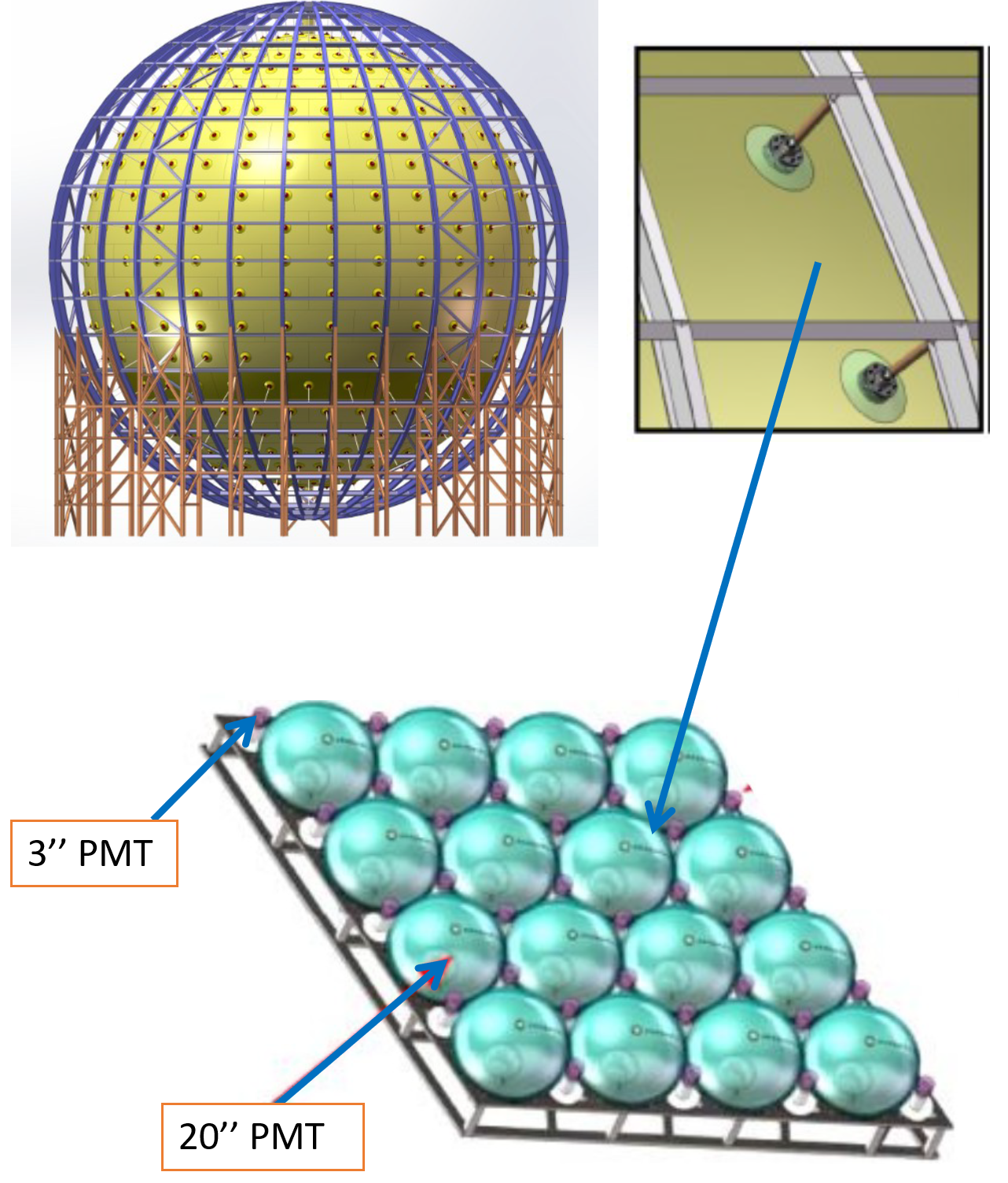}
    \caption{Schematic view of the detector and PMTs installation layout of the JUNO experiment.}
    \label{fig:my_label}
\end{figure}

\hspace{24pt}The energy resolution expected for JUNO is of~3\% at 1 MeV and the energy scale uncertainty should be less than 1\%. To achieve this precision, JUNO was designed as follows.
The detector will be located at $700~m$ underground. JUNO uses 20 ktons of liquid scintillator contained in a acrylic vessel with a 35 m diameter as central detector (CD) \cite{Djurcic}. The acrylic sphere is instrumented by photo-multiplier tubes (PMTs): 17612 20-inch photo-multiplier tubes (LPMT) and 25600 3-inch PMTs (SMPTs) \cite{JUNO_phys}. 
As shown on the right panel of Figure~\ref{fig:my_label}, the SPMTs will be installed in the gap between LPMTs.  \par
As we can see in Figure \ref{fig:my_label}, the central detector is immersed in an ultra-pure water pool of 43.5 m by 44 m which is instrumented by 2400 20-inch PMTs. The pool shields the inner part of the detector from the environment radioactivity and works as a muon veto together with the top tracker on top of the whole structure. The top tracker will study a sample set of muons to reconstruct the muon track and study the background contamination in the CD due to cosmogenic isotope ($^9 Li/ ^8 He$) and the radioactivity from rock. \par
On site, the excavation of the detector cave has been performed, the water-pool concrete structure has been built. The pre-assembly of stainless steel structure has been finished.
All the flat panels of the acrylic sphere have been produced and 80\% of the thermal forming is already performed.

\section{Electronics readout system}
\hspace{24pt}To be able to discriminate between the two NMH, the nominal experimental setup must have an energy resolution of~3\%~at~1~MeV, which corresponds to photon-electron statistics of 1200 p.e./MeV. To achieve this goal with the central detector sizes and specifications, the photo-cathode total coverage of JUNO will be 77\%, and the Photo-detection efficiency  (PDE) greater than 25\%. The PDE of the PMT is the ratio of the detected signal to the input signal of PMT. It is a key factor to achieve the required precision on the energy spectra measurement of the anti-neutrinos.
The photon arrival time measurement precision is lower than 1 ns. This precision is needed for good vertex reconstruction and muon track reconstruction/isolation in the central detector. 
A dynamic range of 1 to 4000 p.e.  (for atm-, geo-, and solar neutrinos) and a trigger with no dead-time (for supernova events lasting up to few seconds) are also required to the readout system.
As a significant part of the electronics will stand underwater, the reliability of the underwater electronics is a main concern. The objective is to have less than 1\% PMT and underwater electronics failure over 6 years of data taking.

\subsection{20-inch PMTs}
\begin{figure}
    \centering
    \includegraphics[width=0.8\textwidth]{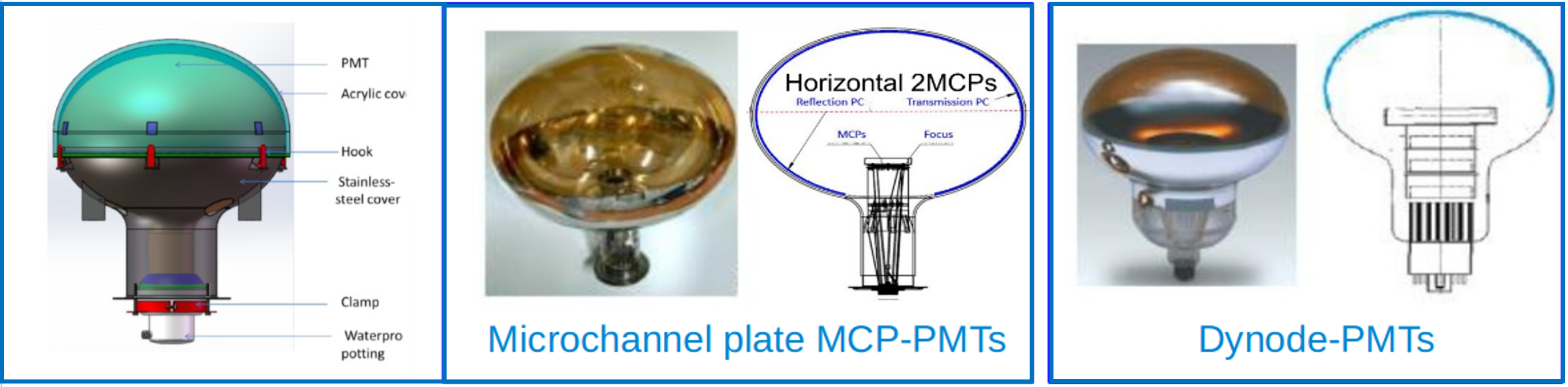}
    \caption{Left panel : PMT mechanical support, central panel : Multi-Channel Plate (MCP) PMT, right panel : dynode-PMT}
    \label{fig:my_label1}
\end{figure}

\hspace{24pt}For the 20-inch photo-multipliers system, JUNO made the choice to use two different technologies: the classical dynode-PMTs (right panel of Figure \ref{fig:my_label1}), and a new  development PMTs (conducted and patented by the IHEP laboratory) based on a Multi-Channel Plate (MCP) PMT (center panel of Figure \ref{fig:my_label1}). More than 20,000 photo-multipliers are produced to instrument the CD and the veto system.  15,000 MCP-PMTs are produced by NNVT and 5,000 dynode-PMTs by Hamamatsu Photonics. The MCP-PMTs are used for the CD and the water pool veto-system while the dynode-PMTs used exclusively for the central detector.
All the PMTs have already been delivered and tested. 
The PDE on all the PMTs achieved is 29.6 \%. More specifically, the mean PDE of 30\% was achieved for, MCP-PMT  and  28.4\% for dynode-PMTs. With the technical improvement from NNVT, the quantum efficiency of MCP-PMT has increased to 35\% (HQE-MCP-PMT) \cite{zhu2018improvement}, the typical quantum efficiency of dynode-PMT is 30\%. The collection efficiency of MCP-PMT is 99.9\%, higher than the one of dynode-PMTs which is 93.3\% \cite{ma2021study}.
Figure \ref{fig:my_label1} left panel shows the protection surrounding the 20-inch PMTs. An acrylic protective cover which reduces the shock wave has been developed. This cover prevents a chain implosion of the tubes in JUNO in case of single PMT implosion, as it happened to Super-Kamiokande experiment in 2001.
\begin{figure}
    \centering
    \includegraphics[width=1.1\textwidth]{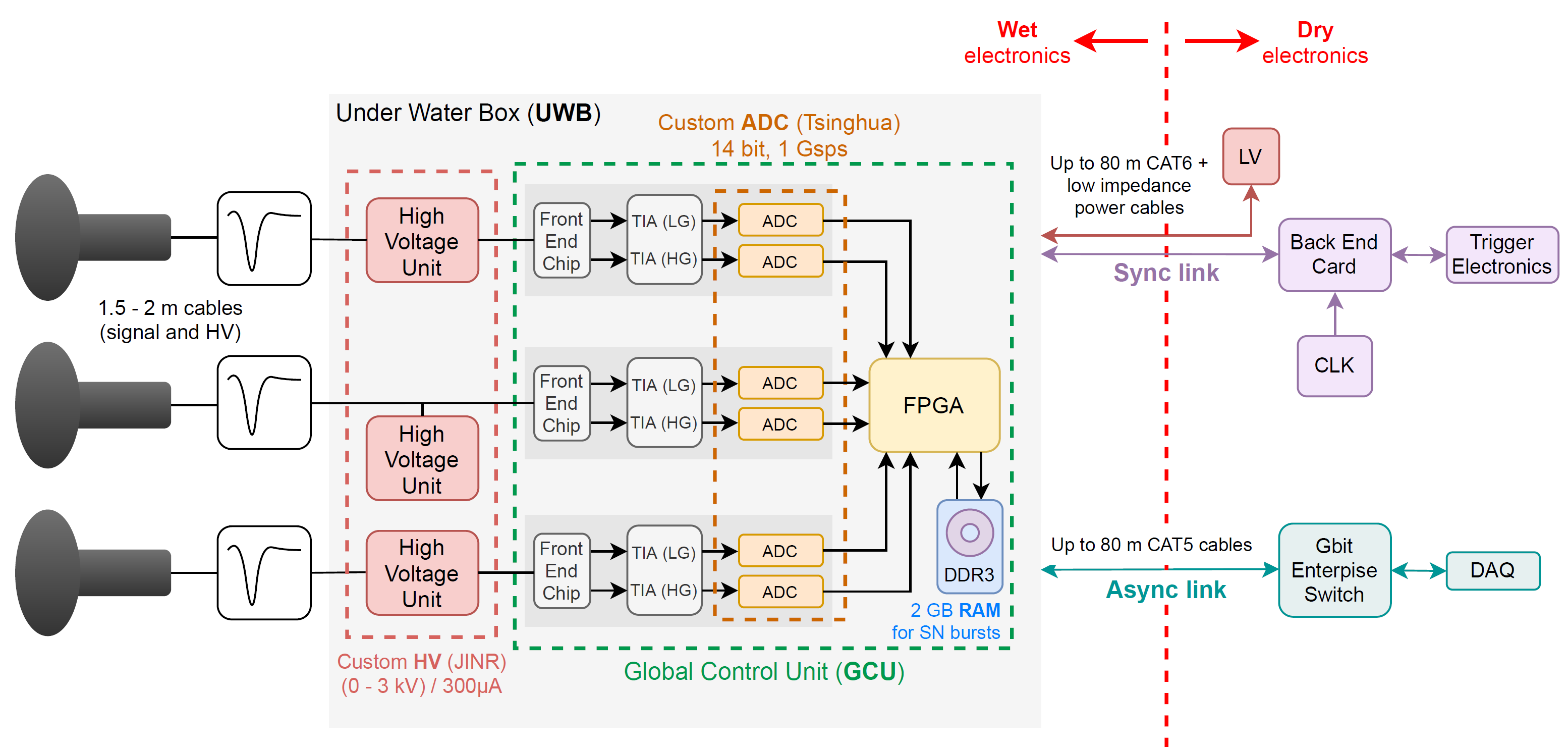}
    \caption{Schematic view of the electronics readout system of JUNO.}
    \label{fig:my_label2}
\end{figure}

\subsection{Readout scheme for the 20-inch PMT}
\hspace{24pt} The JUNO electronics system \cite{Bellato2021} can be separated into two main parts as shown in Figure~\ref{fig:my_label2}.
The left part of Figure \ref{fig:my_label2} shows the underwater electronics. Three PMTs are connected to one underwater box. The PMT housing is connected to the box through a 1 m cable. The box contains the high-voltage unit and the Global Control Units (GCU). The GCUs perform the analog to digital conversion and a first processing of the PMT signal and manage the slow control of the three PMTs. The detailed PMT signal processing chain is the following. The first stage conditions the PMT signal and duplicates it into two streams. The current signal streams are converted to voltage by trans-impedance amplifiers (TIA).
One channel is used for the low gain (8:1) in the case of 0-4000 p.e., and a second channel is used for the high gain (1:1) in the case of 0-128 p.e. . 
A 14-bit 1 Gsample/s custom designed ASIC, developed by Tsinghua University, coverts the analog signals to digital signals. The digital signals are then processed in the Field-Programmable Gate Array to generate trigger request and to store the data into buffer waiting for trigger acknowledgement. A 2 GB DDR3 RAM is used to store the timestamp and charge with the digitized wave-forms. All GCUs, about 7000, are synchronised within less than 10 ns precision. \par %The references \cite{GCU,Timing_Pedretti} explain more extensively the GCUs and the clock system, and the timing synchronisation. \par
\hspace{15pt}The right part of Figure \ref{fig:my_label2} shows the dry electronics. 
The underwater boxes are connected to the dry electronics through two distinctive links : the asynchronous link and the synchronous link. 
The asynchronous link has a variable latency and uses a CAT5E Ethernet link. It is used for the Data readout and slow control of the GCUs. 
The protocol used for this link is IPBUS and works at a nominal speed of 1 Gbps. 
The synchronous link has a fixed latency link. It uses a timing and trigger control (TTC) protocol. 
The nominal link speed is 125 Mbps. Each of the 160 back-end card (BEC) is connected to 44 underwater boxes through CAT6 Ethernet cables. They are used as concentrator cards to collect and compensate the incoming trigger request signals.
\subsection{3-inch PMTs readout system}
\hspace{24pt}The readout system of the 25600 SPMTs is similar to the one of the LPMTs, in particular for the outside water part. 128 SPMTs are connected to an underwater box.
The underwater box includes 3 stages: the high voltage power supplies (same high voltage unit as for the LPMTs), splitting signals, the readout and digitization of the signal. On the ASIC Battery Card (ABC) front-end card, an FPGA controls 8 Charge And Time Integrated Read Out Chip (CATIROC)  ASICs \cite{conforti2021catiroc} that perform the readout and the digitization. CATIROC ASICs are designed to work in single photo-electron mode.
They provide a charge measurement over a dynamic range from 1 to several hundreds of photo-electrons.
The 200 underwater boxes from the SPMTs are connected to the outside water system in the same way as for the LPMTs. \par
As the data of SPMTs have reduced systematics, those data are used to correct the energy reconstruction of the LPMTs data. The SPMT system are also designed to increase the JUNO dynamic range.
In addition, SPMTs improve the muon tracking and give complementary information for supernova studies and for solar oscillation parameter measurements. 

\section{Conclusion}
\hspace{24pt}JUNO will be the largest reactor neutrino detector ever built (20 ktons of LS) with unprecedented energy resolution (3\% at 1 MeV).
 JUNO's main objective is to determine the neutrino mass hierarchy (3-4 $\sigma$ with 6 years of data taking) and four neutrino oscillation parameters to sub-percent level.
JUNO also has a rich physics potential with supernova neutrinos, geo-neutrinos, solar and atmospheric neutrinos.
The production of the different parts and the civil engineering are well underway. The production and assembly of electronic components is going on and the installation will start soon.
The detector assembly completion is expected by the end of 2022 and the data taking is expected to begin in 2023.

\bibliography{biblio}
\bibliographystyle{unsrtnat}
\end{document}